\documentclass[usegraphicx,usenatbib]{mn2e}
\usepackage{enumerate}
\usepackage{graphics}
\usepackage{color}
\usepackage{mathrsfs}
\usepackage{amsmath}
\usepackage{comment}

\newcommand{\msun}{M_\odot}
\newcommand{\rsun}{R_\odot}
\newcommand{\lsun}{L_\odot}

\begin{document}

\title{Mass transfer and magnetic braking in Sco X-1}

\author[Pavlovskii \& Ivanova]{K.~Pavlovskii,$^1$ N.~Ivanova,$^1$ \\
$^1$University of Alberta, Dept.\ of Physics, 11322-89 Ave, Edmonton, AB, T6G
2E7, Canada}

\maketitle
\begin{abstract}
Sco X-1 is a low-mass X-ray binary (LMXB) that has one of the most precisely determined
set of binary parameters such  as the  mass accretion  rate, companions
mass   ratio  and   the  orbital   period.   
For   this  system, as well as for a large fraction of other well-studied LMXBs,  the
observationally-inferred  mass accretion  rate  is  known to  strongly
exceed  the theoretically  expected  mass  transfer rate.  
We suggest that this discrepancy can be solved by applying a modified magnetic 
braking prescription, which accounts for increased wind mass loss in evolved stars 
compared to main sequence stars. 
Using our  mass transfer framework based on {\tt  MESA}, we explore a large 
range of binaries at the onset of the mass transfer.   
We identify  the  subset of binaries  for  which the  mass
transfer tracks cross the Sco X-1 values for the mass  ratio and the
orbital period.   
We confirm that no solution can be found for
which the standard magnetic braking can provide the observed accretion
rates, while  wind-boosted magnetic  braking can provide  the observed
accretion rates for many progenitor binaries that evolve to the observed
orbital period and mass ratio. 
\end{abstract}

\begin{keywords}
binaries: close;
X-rays: binaries;
stars: magnetic field;
methods: numerical

\end{keywords}

\section{Introduction}

Sco X-1 is the the first extrasolar X-ray source discovered and is 
the brightest persistent X-ray source beside the Sun  \citep{1962PhRvL...9..439G}.
The source has been extensively observed since its discovery in both radio and X-ray, 
and is classified as a low-mass X-ray binary (LMXB) for which a number of parameters have been obtained (see Section~\ref{sec:observed}).

Evolution of a close binary after the compact object formation,  
and, specifically, the mass transfer (MT) rate during an LMXB phase, are governed by
angular momentum loss.
In a persistently accreting short-period binary with a giant or subgiant donor, 
the dominant angular momentum loss mechanism is magnetic braking \citep{Rappaport82}.
It is crucial that the first detailed numerical study of the population of short-period LMXBs 
has shown the unexpected strong mismatch between the MT rates in the theoretical population of LMXBs 
and the observationally-inferred mass accretion rates of known LMXBs  \citep{Podsiadlowski02}.
Most of the observed LMXBs, including Sco X-1, were found to have their mass accretion rates  
at least an order of magnitude higher than the theoretically expected MT rates for the same orbital periods
(although it was unclear whether observational selection effects could cause this effect).

This alarming mismatch between  theory and observations, albeit mainly noticed only by theorists doing detailed studies of the MT in LMXBs, 
have prompted several new ideas to explain how the magnetic braking operates in LMXBs.
For example, \cite{Justham06} suggested that Ap and Bp stars with radiative envelopes, due to their high magnetic fields and stellar winds
could nevertheless experience a stronger than usual magnetic braking. 
Another idea, also related to the nature of the donor, 
was that a stronger magnetic braking could be achieved in short-period LMXBs if they have pre-main sequence donors \citep{Ivanova06}.
How the transferred material can affect the orbital evolution also was explored.
For example, \cite{2008A&A...488..257Y} resolved the magnetic braking problem by proposing that accretion discs in LMXBs are truncated,
and \cite{2006MNRAS.373..305C} explored the role of the circumbinary disc.
At the moment, none of the ideas have become the mainstream method for producing LMXBs in population studies,
despite inability of the standard magnetic braking theory to explain the LMXBs that are determined well.
It is important to add that all the mentioned ideas were explored on the set of very short-period LMXBs, with the orbital periods less than about 10 hours, 
and are not applicable to LMXBs with larger orbital period, like Sco X-1.

These facts led us to revisit the way magnetic braking is treated in LMXBs with subgiant and giant donors.
We take the well-observed prototype low-mass X-ray binary -- Sco X-1 -- as an example for our simulations. We provide
its observationally-derived properties in Section 2.
In Section 3, we outline our revised prescription for magnetic braking.
We provide detailed MT simulations for the case of Sco X-1 in Section 4.

\section{Observed parameters of the system}
\label{sec:observed}
From spectroscopic data, \cite{Steeghs2002} infer the upper limit of mass ratio to be 0.46 (see the brief overview of the important observed parameters in Table~\ref{tab:mb_scox1}). 
Assuming that the lowest-velocity emission lines originate at the inner Lagrangian point, 
using the radio inclination data from \cite{Fomalont01} 
and setting the accretor mass to $1.4~M_\odot$ gives a probable value of mass ratio 0.30 and donor mass $0.42 M_\odot$, in agreement with
\cite{Mata15}, who find the donor mass to be between 0.28 and 0.70~$\msun$.
The orbital period of the system is the best determined quantity, and is $P=0.787313\pm0.000015$d \citep{Hynes2012}.

The donor has been identified to have spectral class K4 or later, and its luminosity class IV, subgiant \citep{Mata15}. 
The intrinsic effective temperature of the donor then should not be hotter than 4800~K. This upper limit includes uncertainties in the spectral
class determination as discussed in \cite{2000asqu.book.....C}, for a subgiant the intrinsic effective temperature is likely lower than this upper limit.

The X-ray luminosity of this system,  as measured in 2-20 keV range, 
is $2.3 \times 10^{38}~\rm{erg  s^{-1}}$ \citep{Bradshaw99}.
An estimate of its bolometric luminosity, from the bolometric flux and distance gives $L_{b} =  
3.6\pm0.8 \times 10^{38}~\rm{erg  s^{-1}}$ \citep{Watts08}. 
If the accreting star is $1.4~\msun$, the accreted material is 70\% hydrogen and opacities are provided by Thompson scattering,
then the critical Eddington luminosity for this system is  $L_{\rm Edd, TS}=2.1\times 10^{38}~\rm{erg s^{-1}}$.
This system is slightly above the Thompson scattering limited Eddington limit.
The minimum MT rate that can support this bolometric luminosity for a non-rotating neutron star 
is  $\dot{M}_{\rm min} = 3.5\pm0.8 \times 10^{-8}~\rm{\msun  yr^{-1}}$ 
(for a $1.4~\msun$ neutron star with a radius of 11.5 km), and if one takes into account only
X-ray output, then  $\dot{M}_{\rm min}^{X} \approx 2.2 \times 10^{-8}~\rm{\msun  yr^{-1}}$.

There is an evidence for the presence of a jet originating at the accretor, for details see \cite{Mirabel99}. We have not found any data on the possible circumbinary disk, which
means that the non-conservative mode of MT, in which the specific angular momentum of the lost material is equal to that of the accretor is plausible.
The possible transient nature of Sco X-1 has, to our knowledge, never been discussed, which implies that it is in a steady state,
in which the relationship between accretion rate and X-ray luminosity stays roughly the same with time.

\begin{table}
\caption{Observational properties of Sco X-1}
\begin{tabular}{p{30mm}cc}
\hline
Quantity &  Value & Reference \\
\hline

Mass ratio & $\approx$0.30 but $\la$0.46 & S02 \\
Donor spectral class & later than K4 & M15 \\
Donor luminosity class & IV & M15 \\
Period, d & 0.787313$\pm$0.000015 & H12 \\
Distance, kpc & 2.8$\pm$0.3 & B99 \\
X-ray flux [2-20keV], erg s$^{-1}$ cm$^{-2}$ &  2.4$\cdot10^{-7}$  &  B99 \\
Bolometric flux, erg s$^{-1}$ cm$^{-2}$ &  3.88$\cdot10^{-7}$  &  W08 \\
\hline
\end{tabular}
\label{tab:mb_scox1}
\medskip

Sources: S02 --  \cite{Steeghs2002}, H12 --  \cite{Hynes2012}, B99 --  \cite{Bradshaw99}, W08 -- \cite{Watts08},
M15 -- \cite{Mata15}.
\end{table}

\section{Magnetic braking}
\label{sec:mb}

It is widely accepted that magnetic braking  -- the removal of the
angular momentum from a rotating star  by the action of a magnetically
coupled stellar  wind -- is crucial  for studies of the  formation and
evolution of a  number of classes of close binaries.  It has been most
deeply  addressed for  the  case of  cataclysmic variables  \citep[for
  details, see][]{2011ApJS..194...28K}, but it  also plays an important
role for LMXBs.

It was \cite{1962AnAp...25...18S} who recognized  first that  the slowing
down of  single stars can take  place when the material  lost from the
stellar surface  is kept in corotation  with the star by  the magnetic
field. As a result, the specific  angular momentum carried by the gas
is significantly greater than in a spherically symmetric stellar wind.
The  corotation  can be  achieved  if  a  star possesses  a  substantial
magnetic field.  The strength of the magnetic filed has been linked to the
generation of a  magnetic field by dynamo action in  a deep convective
envelope.

\cite{Skumanich72} had  found observationally  that, in  main sequence 
stars of spectral class G, the equatorial rotation velocities decrease
with time,  $t$, as  $t^{-0.5}$.  This  timescale of  angular momentum
removal can  take place if  the rate of  the angular momentum  loss, $\dot
J_{\rm  MB}$, is  proportional to  $\Omega^3$, where  $\Omega$ is  the
stellar rotational velocity.  
After being calibrated to the  observed angular momentum losses in main
sequence  stars in  open clusters,  this  law is  usually referred to as the
Skumanich magnetic braking law,  and can be written in a  generic form as considered
by \cite{Rappaport83}:

\begin{equation}
\dot J_{\rm{mb,Sk}} = -3.8 \cdot 10^{-30} M \rsun^{4} \left(\frac{R}{\rsun} \right)^{\gamma} \Omega^{3}~\rm{dyne~cm} \ .
\label{eq:skum}
\end{equation}

\noindent Here  $M$ and $R$  are the mass and radius  of the star  that is
losing its angular  momentum via magnetic braking, and $\gamma$  is a dimensionless
parameter from 0  to 4.  The magnetic braking is observationally  absent in low-mass
main sequence  stars; this was linked  to the possible halt of  the dynamo
mechanism in  almost fully convective  main sequence stars.   For fast
rotators, it  has been  discussed that the  dipole magnetic  field can
create a dead zone that traps  the gas, or alternatively that once $B$
has reached some maximum value, it  saturates and can not increase any
further
\citep[e.g.][]{1987MNRAS.226...57M,2003ApJ...599..516I,2003ApJ...582..358A}.
In  either case,  $\dot  J_{\rm  MB}$ has  a  shallower dependence  on
$\Omega$,  approaching $\Omega^2$,  and its value  is smaller  than
the Skumanich law would predict.

Let us estimate what MT rate the Skumanich magnetic braking provides. 
For a binary system, assuming a conservative MT,

one can find that the accretion rate is


\begin{align*}
\frac{\dot{M}}{M} = - \frac{\dot{J_{\rm MB}}}{J_{\rm orb}} \frac{1}{0.8(3) - q},
\end{align*}
where q is the mass ratio (donor to accretor). This formula is obtained using the Roche lobe radius approximation by \citep{1971ARA&A...9..183P}, specifically,  that for a mass ratio below about 0.5, the ratio of the Roche lobe radius and binary separation is approximately $0.46224 (1 / (1 + q))^{1/3}$. 

When the mass ratio $q = 0.3$, it gives approximately 
\begin{equation}
\label{eq:mdotm}
\frac{\dot{M}}{M} \approx -2\frac{\dot{J_{\rm MB}}}{J_{\rm orb}}   
\end{equation}
\noindent and 
\begin{eqnarray}
\dot M &=& 6.1 \cdot 10^{-9}  M_\odot {\rm yr}^{-1} \times \\ & &\left(\frac{R}{\rsun} \right)^{\gamma}\left(\frac{a}{\rsun} \right)^{-2} \left ( \frac{1 {\rm day}}{P_{\rm orb}} \right )^2 \frac{M+M_2}{M_2}\ \frac{M}{M_\odot}\ . \nonumber 
\end{eqnarray}

\noindent  Here $M_2$  is the  mass of  the binary  companion.  For  a
system of a 0.4~$\msun$ subgiant and a 1.4~$\msun$ neutron star at the
observed period, the Skumanich prescription  with $\gamma$ from 0 to 4,
${\dot  M}  = 2.6$ to $6\times  10^{-10}~\rm{\msun  yr^{-1}}$.
Clearly, as  was found  in the previous  studies, the  Skumanich law
provides the  MT rate that is two orders of  magnitude lower than
the observed mass accretion rate, and is independent of how we evolve the star.

The mass ratio and donor mass in Sco X-1 are not very certain.
The recent paper by  \cite{Mata15} lists
the intervals for the mass of the donor from 0.28 to 0.7~$\msun$ and
for the mass ratio from 0.28 to 0.51. We can test the
effect of this uncertainty on our estimate.
We considered the combinations of mass ratios 0.28, 0.5, 0.6, 0.7
and donor masses 0.28, 0.4, 0.51~$\msun$. In all considered cases the mass accretion
rate following from the Skumanich law doesn't 
exceed $2.7\times  10^{-9}~\rm{\msun  yr^{-1}}$. This ensures that
the uncertainties in the mass ratio and the donor mass are not likely
to be the reason for the discrepancy between the observed
mass accretion rate and the theoretical estimate.

We note however that  Sco X-1 is an evolved star. In  this case, it is
important to realize  that Skumanich magnetic braking law was {\it  scaled} to match
the observations of main sequence  stars. This implicitly included two
assumptions: (i) the  wind mass loss is as at the main sequence rate, and (ii)
the  magnetic  field  strength  is  only  changing  with  the  angular
velocity.

Let us  consider how these two  quantities can affect the  rate of 
angular  momentum  loss via magnetic braking.   From  continuity, and  assuming  an
isotropic wind;

\begin{equation}
\label{eq:mass_flux}
\dot M_{\rm w} =  4 \pi R_{\rm A}^2 \rho_{\rm A} v_{\rm A} =  4 \pi  R^2 \rho_{\rm S} v_{\rm S}\ .
\end{equation}

\noindent  Here $R_{\rm  A}$  is  the radius  of  the Alfven  surface,
$\rho_{\rm A}$  and $v_{\rm A}$  are the density  and the speed  of the
material that  crosses it; $R$ is  the radius of the  star, $\rho_{\rm
  S}$ and  $v_{\rm S}$  are density  and velocity of  the wind  at the
star's surface. The  angular momentum loss through  the Alfven surface
is then

\begin{eqnarray}
\label{eq:torque}
\dot J_{\rm MB} & = & 
 - 4 \pi \Omega \int_{0}^{\pi/2} \rho_{\rm A} v_{\rm A} R_{\rm A}^2  \left (R_{\rm A}  \sin \theta \right )^2 \sin \theta d\theta \\ &\simeq&
- \frac{2}{3} \dot M_{\rm w} \Omega  R_{\rm A}^2 \nonumber 
\end{eqnarray}

\noindent Here  we assumed  for simplicity that  $R_{\rm A}$  does not
depend on $\theta$, which is not necessarily true.  The Alfven surface
is defined  as the surface  where the  wind speed $v_{\rm  w}$ becomes
the Alfvenic speed,  $v_A$, or in  other words, the magnetic  pressure and
ram                pressure                are                balanced
\citep[e.g.,][]{1968MNRAS.138..359M,1987MNRAS.226...57M}

\begin{equation}
\frac{1}{2} \rho_{\rm A} v_{\rm A}^2 \simeq  \frac{B(r)^2}{{8\pi}}\ .
\label{eq:alfv}
\end{equation}

\noindent In the case of a radial magnetic field $B(r)=B_{\rm S} R^2/r^2$,
and in a case of a  dipole field, $B(r)=B_{\rm S}R^3/r^3$, where $B_S$
is the surface magnetic field. For a radial field,

\begin{equation}
\label{eq:radial_field}
\frac{1}{2} \rho_{\rm A} v_{\rm A}^2 \simeq  \frac{B_{\rm S}^2}{{8\pi}} \frac{R^4} {R_{\rm A}^4} \ .
\end{equation}

\noindent  To  close the  system,  one  more important  assumption  is
needed, about  the wind velocity  along the magnetic  streamlines, and
this  is where most  of the uncertainty  is hidden.   The most  often
considered option is to assume that  the system is isothermal. Then one
can consider the generalized Bernoulli  equation for a rotating system
inside    the    corotating     zone    \citep[e.g.,    Equation    A8
  in][]{1987MNRAS.226...57M} .
In this case, it  can be shown that velocity at  the Alfven surface is
reduced  to  the wind sonic  velocity  $c_{w}$\citep{1987MNRAS.226...57M}.
By combining equations (\ref{eq:mass_flux}) and (\ref{eq:radial_field}) and further assuming that  $v_{\rm A} = c_{\rm w}$, we get

\begin{equation}
R_{\rm A}  \simeq  B_{\rm S} \frac{R^2} {\sqrt{\dot M_{\rm w}  c_{\rm w}}} \ .
\end{equation}

\noindent  With the  further standard  assumption of  $B_{\rm S}=  B_0
\Omega $ (later in this Section  we will show where this assumption 
comes  from) we  recover Equation \ref{eq:torque} in the same functional  dependence as  in  the
empirical Skumanich law (see Equation \ref{eq:skum}):
 
\begin{equation}
\dot J_{\rm MB} \propto  B_{\rm S}^2 \Omega R^4  \propto B_{\rm 0}^2 \Omega^3  R^4 \ . 
\end{equation}

In this, and other demonstrated below functional dependencies for the angular momentum loss we show 
power-laws for the most important quantities. As such we keep the quantities
in which $\dot J_{\rm mb}$ is usually expressed in literature -- 
the  magnetic field strength, the angular velocity and the radius of the stars. We also will add
below $\dot M_{\rm w}$. We drop for clarity less important terms that may enter the exact equation  
(e.g., sonic velocity, surface density or mass).

For a dipole field and a similar thermally-driven wind, one can similarly obtain

\begin{equation}
 {R_{\rm A}}   \simeq  B_{\rm S}^{1/2}  \frac{R^{3/2}} {({v_{\rm A} \dot M_{\rm w}})^{1/4}} .
\end{equation}

\noindent Using an additional assumption that the isothermal  wind is of order of the
surface   escape   velocity   when it reaches  the   Alfven   surface
\citep{Justham06} leads to 

\begin{equation}
\label{eq:RA_dipole}
 {R_{\rm A}}   \simeq  B_{\rm S}^{1/2}  \frac{R^{13/8}} {({\sqrt{2 GM} \dot M_{\rm w}})^{1/4}} \ . 
\end{equation}

\noindent Combining Equations~(\ref{eq:torque}) and (\ref{eq:RA_dipole}), we obtain
a functional dependence of the form \citep{Justham06}

\begin{equation}
\dot J_{\rm MB} \propto \dot M_{\rm w}^{1/2} B_{\rm S} R^{13/4} \Omega \propto \dot M_{\rm w}^{1/2}  R^{13/4} \Omega^2 \ .
\end{equation}

\noindent Note that here the wind loss rate enters the functional dependence. 

If one  assumes that  the Bernoulli equation  is legitimate,  then the
other two limiting  cases are that velocity at the  Alfven surface can
be reduced  either to the  local escape velocity,  or is of
order $\sim  \Omega R_{\rm  A}$.  These assumptions  will lead  to
different powers of $\Omega$ and stellar wind mass loss rate in
the magnetic braking torque, thus there is no unique  way to obtain $\dot
J_{\rm MB}$.  This  uncertainty is also reflected in  the existence of
several prescriptions for the magnetic braking law, and  the use of a free value for
the parameter $\gamma$ in the prescription of Equation~\ref{eq:skum}.

Compared to main  sequence stars, subgiants are  generally colder, and
hence the  assumption of isothermal wind  velocity inside  the Alfven
sphere may not  hold -- not even considering that the  wind can be accelerated
by the  magnetic fields.  Indeed, even  for the Sun, we  know that the
temperature  of the  wind  is substantially  higher  than the  surface
temperature,  and  also that  the  wind  is both  heated  and
accelerated             via              several             mechanism
\citep[e.g.,][]{2007ApJS..171..520C}.   Therefore,   the  generalized,
albeit convenient, Bernoulli equation as in \cite{1987MNRAS.226...57M}
is not valid.

Similarly to  the Sun, it was  found in MHD simulations  of red giants
winds that they are  also accelerated \citep{2007ApJ...659.1592S}.  
Unlike the Sun,  winds are also structured, can form bubbles,
and hence a method that assumes existence of the Alfven sphere might be not applicable.
We can nonetheless examine what self-consistent
winds \citep{2007ApJ...659.1592S} in a radial field may imply, by 
considering their
red giant models that are closest to the case of Sco X-1 donor, models
II and III. Importantly, in their models, density drops  with distance as $\propto R^{-3}$.
As a result,  from continuity, velocity grows  linearly with distance.
At  the Alfven surface, the velocity  of the wind may  reach a value
that is about the surface escape velocity \citep[we recognize that this not necessarily the case
for larger giants where the winds are slower than surface escape velocity, see also discussion in][]{2011ApJ...741...54C}.  
Then $R_{\rm A}$ can be expressed only using the surface values of the star,
with the direct dependence on $\dot M_{\rm w}$ disappearing. Assuming that $\rho_{\rm A} = \rho_{\rm S} R^3 / R_{\rm A}^3$ and that $v_{\rm A}^2 = 2 G M / R $, we obtain from Equation~(\ref{eq:radial_field}):

\begin{equation}
\label{eq:ra}
R_{\rm A} = \frac{B^{2}_{\rm S}}{8 \pi} \frac{R^2}{GM\rho_{\rm S}}\ , \
\end{equation}

\noindent where $\rho_{\rm S}$ is linked to the density at the wind base 
and is determined by hydrostatic equilibrium
in stellar photosphere \citep{2007ApJ...659.1592S}. 
By substituting Equation~(\ref{eq:ra}) into (\ref{eq:torque}), we arrive at the functional form for the rate of loss of the
angular momentum in a red giant as

\begin{equation}
\dot J_{\rm MB} \propto \dot M_{\rm w} \Omega B_{\rm S}^4 R^4 \ .
\label{eq:jmb_boosted}
\end{equation}

\noindent Note that  here the dependence on the wind mass loss rate does not
disappear, unlike the standard case that is the isothermal solution in
a radial field. Recall that this standard case is the basis
of the commonly used Skumanich law in the form by \cite{Rappaport83}.

 A similar consideration of a dipole field can be done by supplanting $B(r)$ with $B_{\rm S} (R / r)^3$ in Equation~(\ref{eq:alfv}). Then Equations~(\ref{eq:ra}) and (\ref{eq:torque}) would produce
an angular momentum loss rate that is  proportional to $\dot M_{\rm wind} \Omega B_{\rm S}^{4/3} R^{8/3}$.
As one can see, varying the geometry of the magnetic field changes the power law with which the magnetic field
enters in the functional form. But it is the assumptions on reaching the Alfven surface
for the stars' surface escape velocity and on the density profile that keep the functional 
form proportional to the wind mass loss rate. 

Now we address  the surface value of the magnetic  field.  It has been
discussed in the past that the  dynamo activity scales with the dynamo
number                                                           $N_D$
\citep[e.g.,][]{Parker71_dynamo,Hinata89_dynamo,Meunier97_dynamo}.
The dynamo  number is related to  the Rossby number as  $N_{\rm D}\sim
Ro^{-2}$,  where  Rossby   number  is  defined  as   $Ro  =  1/(\Omega
\tau_{\rm conv})$,  where   $\tau_{\rm conv}$  is   the  convective   turnover  time
\citep{Noyes84_dynamo}.   \cite{Ivanova06}   has  discussed
that in LMXBs with donors that are not on  the main sequence, this has to be taken
into account as

\begin{equation}
B_{\rm S} = B_{\rm S}^{\rm 0} \frac{\tau_{\rm conv}}{\tau_{\rm conv}^{\rm 0}} \frac{\Omega}{\Omega^{\rm 0}} \ .
\end{equation}
\noindent Here indexes ``0'' are for  values of some star with respect
to which the magnetic braking law should be calibrated.

Considering the two factors discussed above,   we propose to examine the
angular momentum loss rate that  is equivalent to consideration of the
Skumanich law in Equation~(\ref{eq:skum})  but with
two additional scaling (``boost'') factors, wind-boost and $\tau$-boost:

\begin{equation}
\dot J_{\rm MB} = \frac{\dot M_{\rm w}}{\dot M_{\sun}} \left ( \frac{\tau_{\rm conv}}{\tau_{\rm conv}^{\sun}} \right )^{\eta}   \dot J_{\rm MB, Sk} \ .
\label{eq:mbboost}
\end{equation}

\noindent The power  $\eta$ with which the $\tau$-boost enters in the equation can vary, it is two for the empirical Skumanich law and 
can be as high as four in the case of a giant wind as discussed above (see Equation \ref{eq:jmb_boosted}).

To account for the wind-boost, for the solar wind loss rate we take ${\dot M}_{\odot} = 2.5
\cdot 10^{-14}$  $M_\odot$ per  year \citep{Carroll95}.   For subgiant  and giant
wind mass loss  rate, we adopt the standard  Reimers wind prescription
\citep{Reimers75}:

\begin{equation}
\label{eq:reimers}
{\dot M}_{\rm{Reim}} = 4 \cdot 10^{-13} \frac{R}{\rsun} \frac{L}{\lsun} \frac{\msun}{M}~\rm{\msun yr^{-1}} \ , 
\end{equation} 
\noindent where $L$ is donor's luminosity.  The values of the expected
boost  in   an  unperturbed  $1\   M_\odot$  giant  are   provided  in
Table~\ref{tab:mbboost}.  We estimated the accretion rate for
the magnetic braking prescription as in Equation \ref{eq:mbboost}, using Equations~(\ref{eq:mdotm}) and (\ref{eq:reimers}). In order to obtain
$L$, that enters Equation~(\ref{eq:reimers}), we  took into  account  the  donor's observed effective  temperature $T_{\rm eff}$ to be 4800~K according to the spectral class given in Table~\ref{tab:mb_scox1}. We assumed that the donor's luminosity
is proportional to its surface area and $T_{\rm eff}^4$.  If the
 boost by the convective turnover time (which we can only
find when  a proper giant model  with lost mass will  be obtained) is neglected, we
obtain $0.74$ to $1.7  \times 10^{-8}~\rm{\msun  yr^{-1}}$, which is close to the observed range for Sco X-1 (see Table~\ref{tab:mb_scox1}).

\begin{table}
\caption{Magnetic braking in an unperturbed 1~$\msun$ star}
\begin{tabular}{ccc}
\hline
$\log_{10}(R / \rsun)$ & ${\dot M}_{\rm Reim} / {\dot M}_{\sun}$ &  $\tau_{\rm conv} / \tau^{\sun}_{\rm conv}$ \\
\hline

0.27 & 4      & 3.7 \\
0.95 & 260    & 9.2 \\
1.23 & 1413   & 9.9 \\
1.43 & 4500   & 11.0 \\
1.57 & 10325  & 11.8 \\

\hline
\end{tabular}
\label{tab:mbboost}
\medskip

$\tau_{\rm conv}$ -- convective turnover timescale in the \texttt{MESA} model. Wind is calculated as in \cite{Reimers75}.
\end{table}

\section{Detailed evolution and mass transfer}

For the simulation of binary MT through
the inner Lagrangian point ($L_{1}$), we use our framework \citep{PavlovskiiIvanova15} based on the {\tt MESA} 
code\footnote{Modules for Experiments in Stellar Astrophysics, http://mesa.sourceforge.net}.
\texttt{MESA} is a modern set of stellar libraries described in \cite{Paxton11, Paxton13}. 
We obtain the binary evolutionary tracks for systems with varying
initial parameters. The donor mass at ZAMS was varied from 0.9 to 1.8~$\msun$ and the initial neutron star mass was varied from 1.24 to 1.6~$\msun$. 
For a fixed combination of masses from these ranges we adjust the initial period to find the tracks that 
pass as close as possible to the point of $q = 0.30$, $P = 0.787$, which corresponds to
the observed parameters of Sco X-1. At the initial orbital period the donor is a ZAMS star orbiting a NS. We use solar metallicity for the donor.

\begin{figure}
\includegraphics[width=90mm]{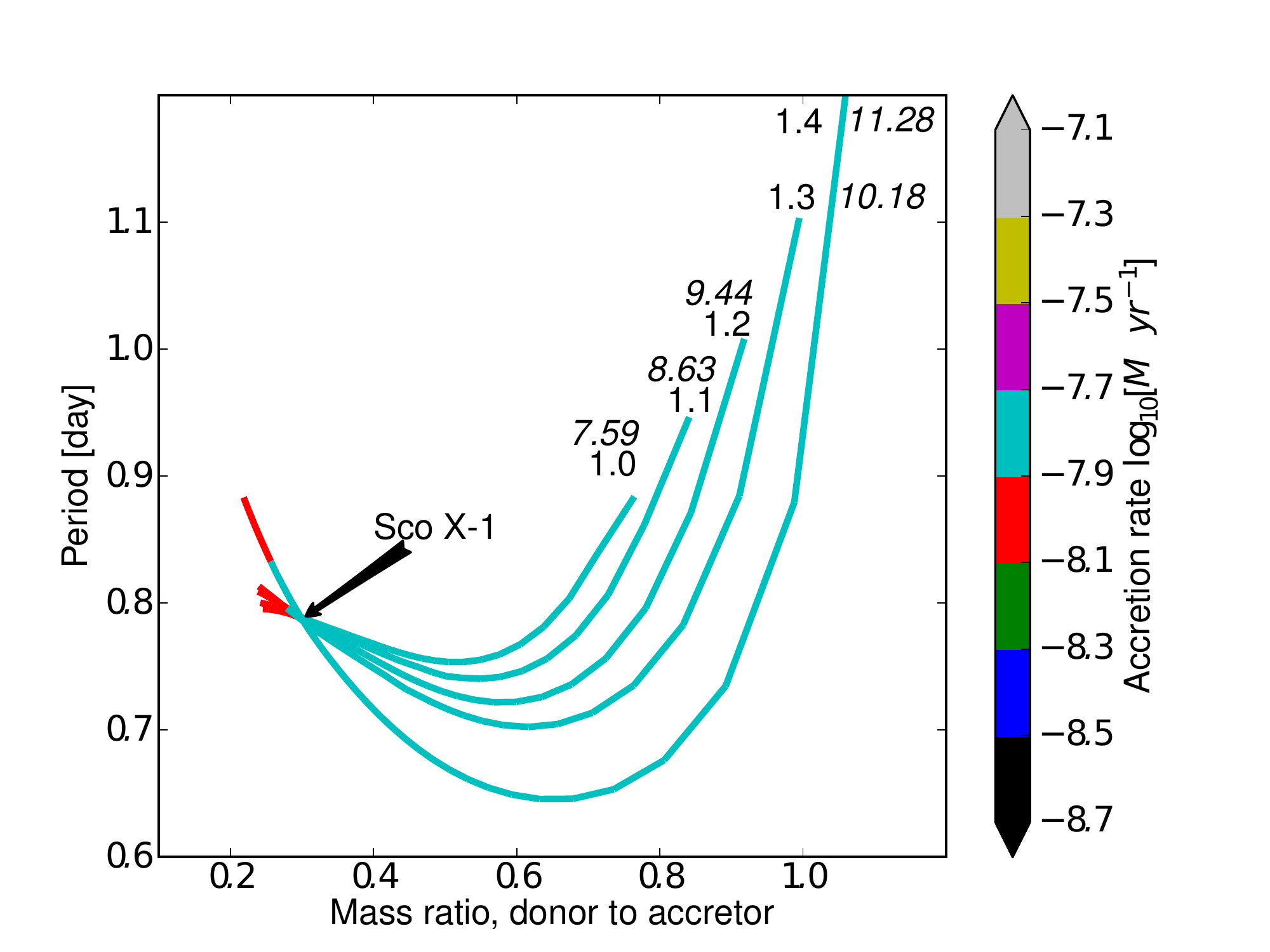}
\caption{Models employing the "wind-boosted" (not "$\tau$-boosted") magnetic braking law, given by Equation~\ref{eq:mbboost}. 
Each track corresponds to the evolution of a binary system that crosses the point corresponding to the observed parameters of Sco X-1, i.e. $q = 0.30$, $P = 0.787$.
Tracks start when more than 0.01~$\msun$ is accreted. Roman numbers denote initial donor mass in solar masses, italic numbers denote initial period in days. The initial NS (accretor) mass for all tracks is 1.3~$\msun$.
Color denotes current mass accretion rate. Note that these simulated mass accretion rates agree with observations.}
\label{fig:tracks_boost_1.3}
\end{figure}

We assume that the compact companion was already a neutron star when the initially less massive 
star overfilled its Roche lobe for the first time at the subgiant stage. We assume that
the accretion rate is Thompson-scattering limited to ${\dot M}_{\rm Edd, TS} = 4 \pi c R / (0.2 (1 + X))$, 
where R is neutron star radius, taken to be 11.5~km, $X$ is the hydrogen abundance in the outer layers of the donor. 
The excess material above  ${\dot M}_{\rm Edd, TS}$ is assumed to be taken away from the system carrying the specific angular momentum of the accretor.

We also assume that the components are circularized at all times, and that magnetic braking torque applied to the donor brakes the whole
system via the tidal interaction. We utilize both the standard and wind-boosted prescription for the magnetic braking.

\begin{figure}
\includegraphics[width=90mm]{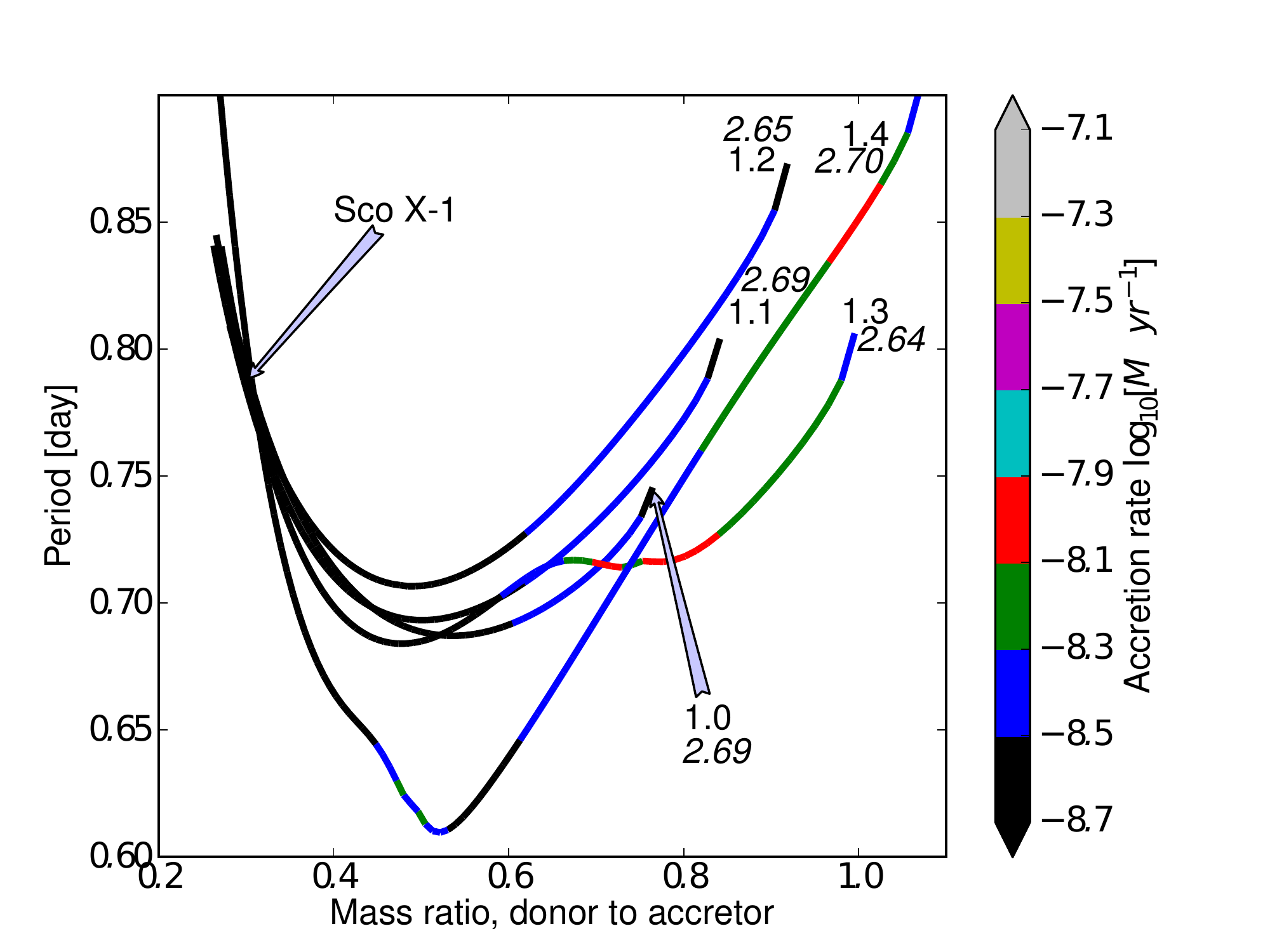}
\caption{Same as in Figure~\ref{fig:tracks_boost_1.3}, but employing the conventional magnetic braking law, given by Equation~\ref{eq:skum} with $\gamma = 3$. Other notations as in Figure~\ref{fig:tracks_boost_1.3}.
Note that the simulated mass accretion rates disagree with observations.}
\label{fig:tracks_noboost_1.3}
\end{figure}

\begin{figure}
\includegraphics[width=90mm]{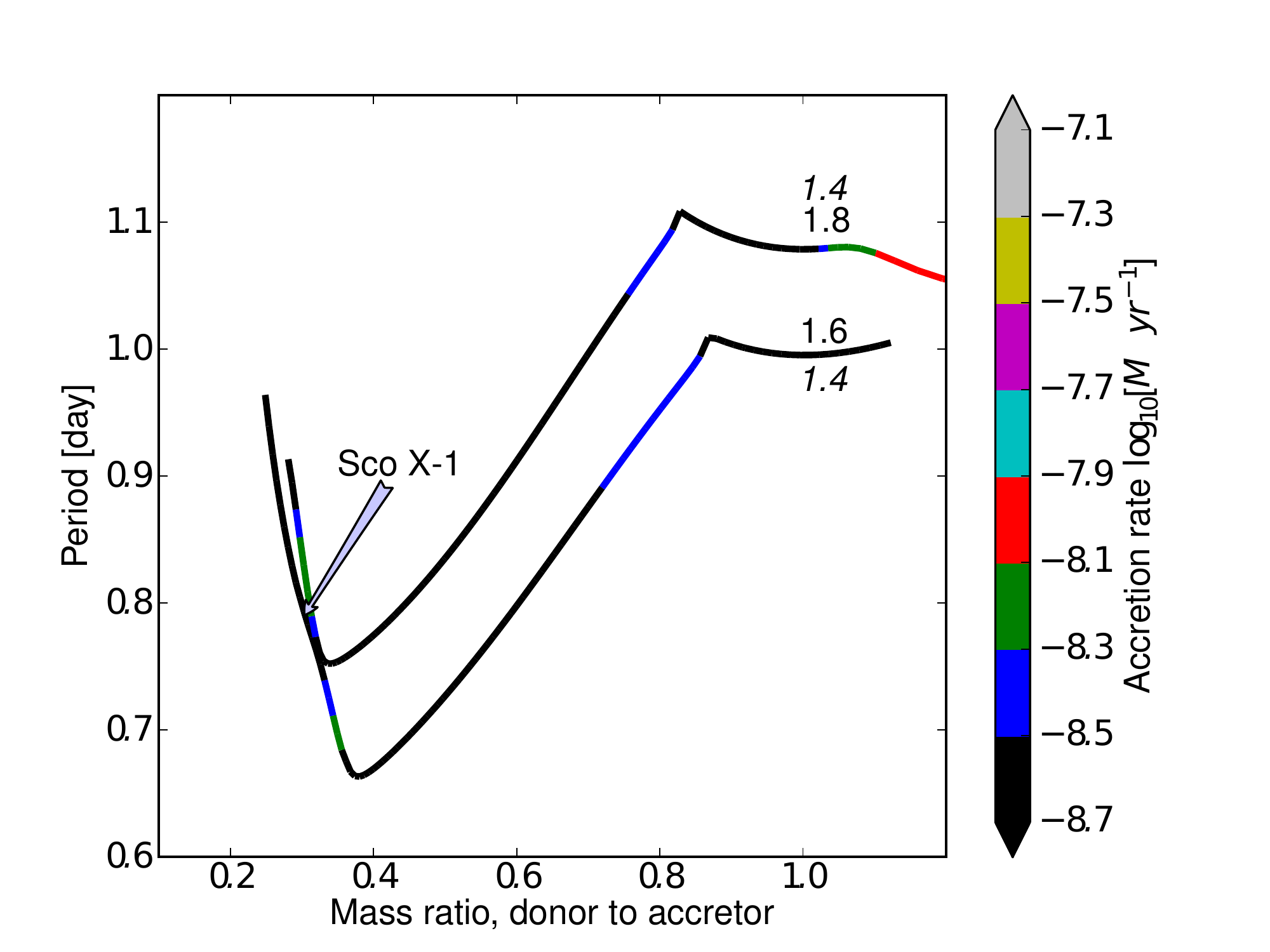}
\caption{Same as in Figure~\ref{fig:tracks_noboost_1.3}, but for initially more massive donors and initial NS mass 1.42~$\msun$. Other notations as in Figure~\ref{fig:tracks_boost_1.3}.
Note that in this case the donor overfills its Roche lobe before the  convective envelope forms.
The simulated mass accretion rates are still too low.}
\label{fig:tracks_massive_noboost_1.42}
\end{figure}

The results of these simulations for both wind-boosted (not $\tau$-boosted) and regular magnetic braking are shown in
Figures~\ref{fig:tracks_boost_1.3}, \ref{fig:tracks_noboost_1.3} and \ref{fig:tracks_massive_noboost_1.42}.
As can be seen in Figure~\ref{fig:tracks_boost_1.3}, in the wind-boosted model the accretion rate at the Sco X-1
point becomes closer to the observational estimate obtained in Equation~(\ref{sec:observed}) as the donor's initial mass increases: the rates are respectively
1.4, 1.5, 1.5, 1.5, 1.8, 2.0 $\times 10^{-8}~\rm{\msun yr^{-1}}$, whereas the observed rate is 2.2 $\times 10^{-8}~\rm{\msun  yr^{-1}}$.
Note also that the obtained value of accretion rate is close to the estimate obtained in Section~\ref{sec:mb}.
For the models utilizing the regular magnetic braking prescription (Fig. \ref{fig:tracks_noboost_1.3}), 
accretion rates vary from 0.1 to 0.2 $\times 10^{-8}~\rm{\msun yr^{-1}}$, which
is more than an order of magnitude less than that observed.
 
\begin{figure}
\includegraphics[width=90mm]{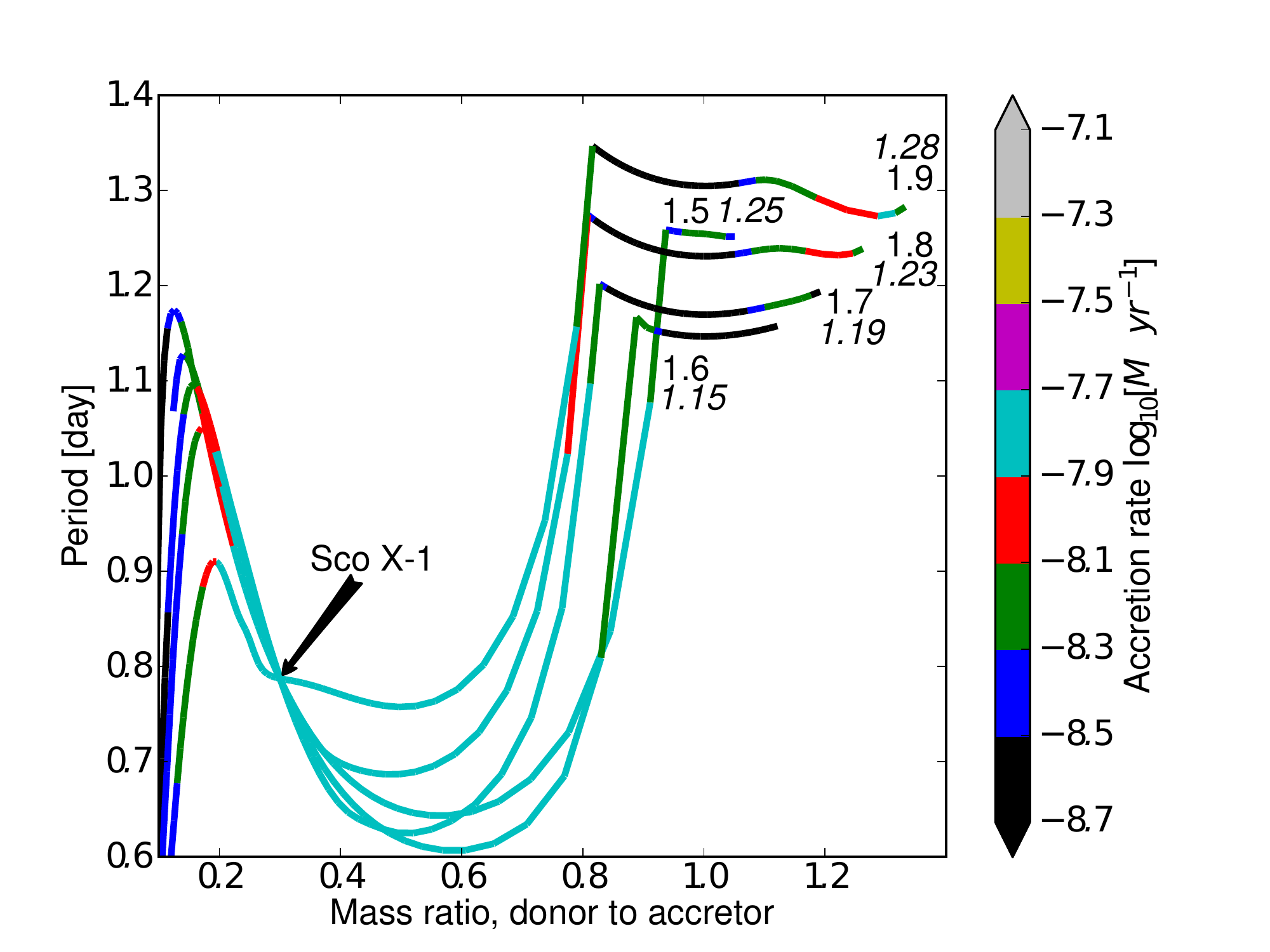}
\caption{Same as in Figure~\ref{fig:tracks_boost_1.3}, but for initially more massive donors and initial NS mass 1.42~$\msun$. Other notations as in Figure~\ref{fig:tracks_boost_1.3}.
Note that in this case the donor overfills its Roche lobe before the  convective envelope forms.
The simulated mass accretion rates agree with observations.}
\label{fig:tracks_massive_boost_1.42}
\end{figure}

Varying the initial mass of the NS and taking more massive donors does change the shape of the tracks that pass through the Sco X-1 point, but the same feature remains unchanged:
wind-boosted tracks have mass accretion rates at the Sco X-1 point comparable to the observations -- for all tracks the mass accretion rate at Sco X-1 point is 2.0 $\times 10^{-8}~\rm{\msun yr^{-1}}$ and
the unboosted tracks lack this agreement with the maximum accretion rate of 0.4 $\times 10^{-8}~\rm{\msun yr^{-1}}$ (see Figures~\ref{fig:tracks_massive_noboost_1.42} and \ref{fig:tracks_massive_boost_1.42}).


Note that the shape of the tracks in Figures~\ref{fig:tracks_massive_noboost_1.42} and \ref{fig:tracks_massive_boost_1.42} is
different from the ones in  Figures~\ref{fig:tracks_boost_1.3} and \ref{fig:tracks_noboost_1.3}.
This is because in this case the donors overfill their Roche
lobes before the deep enough convective envelope develops for the magnetic braking
mechanism to start to operate. A convective envelope is established
only after some mass is lost, this is when the
magnetic braking switches on. These moments correspond
to the turning points on the tracks. In attempt to verify if there are solutions
resembling those in Figure~\ref{fig:tracks_boost_1.3}, without the turning
point and with longer initial periods, we looked at the initial
periods from 1.5 to 20 days for the 1.5~$\msun$ donor and found no such solutions.

We find that systems with a more massive initial NS mass and donor mass experience non-conservative MT at the Sco X-1 point.
For these systems, in the order of increasing donor ZAMS mass from 1.5~$\msun$ to 1.9~$\msun$,  
the mass transfer rates are 2.3, 2.7, 3.3, 3.3, 2.5~$\times 10^{-8}~\rm{\msun yr^{-1}}$ (see  Figure \ref{fig:tracks_massive_boost_1.42}).   
Among the tracks with initially 1.3~$\msun$ accretor and 1.0 to 1.5~$\msun$ donor ZAMS mass, only the one obtained with 1.5~$\msun$ donor ZAMS mass is non-conservative with mass transfer rate
2.2~$\times 10^{-8}~\rm{\msun yr^{-1}}$ (this track is not shown in Figure~\ref{fig:tracks_boost_1.3}). All tracks obtained with the classical magnetic braking prescription have conservative mass transfer at the Sco X-1 point.

The effective temperature of the donor is 4661~K for 1.0~$\msun$ donor and 1.3~$\msun$ NS, and increases further with both initial donor mass and NS mass.
For example, a 1.1~$\msun$ donor with a 1.3~$\msun$ NS would already have $T_{\rm eff}=4692$~K 
and a 1.0~$\msun$ donor with a 1.42~$\msun$ NS has $T_{\rm eff}=4710$~K.
Because the maximum effective temperature of the donor is 4800~K, systems with donor ZAMS mass  $\ga1.6$~$\msun$ are unlikely to be the progenitors
because in this case even for a 1.3$\msun$ neutron star the resulting effective temperature of the donor exceeds the 4800~K limit.
If we fix the mass of the neutron star at 1.42~$\msun$, then based on the observed accretion rate, the most likely ZAMS mass of the donor is between 1.4 and 1.5~$\msun$.

Finally one can estimate to which degree the difference in convective turnover timescale could affect these results.
For this estimate one needs to know how the convective turnover time of the donor at the Sco X-1 point relates to the solar convective turnover time.
In our models the ratio $\tau_{\rm conv} / \tau^{\sun}_{\rm conv}$ at the Sco X-1 point reaches 4.
The magnetic braking torque could be additionally boosted by this factor, however this will not affect 
the mass accretion rates in those models, which already accrete at the Eddington rate, e.g. those shown
in Figure~\ref{fig:tracks_massive_boost_1.42}. We note that without any wind boost, and only considering the boost of the magnetic field due to convective turnover,  
the observed accretion rates can also be achieved, but we do not have detailed tracks for this case. 
We note that in the case when wind-boost is not taken into account, $\eta$ should be taken as 2 in Equation~\ref{eq:mbboost}.

Using the wind-boosted magnetic braking law, for every considered combination of the initial masses of 
the donor and compact object we were able to find the initial period
that led the binary system to the observed period and mass ratio of Sco X-1 and to the accretion rate comparable to that of Sco X-1.
Similar to the results discussed in \cite{2005A&A...440..973V}, we see that 
the evolution of the orbital periods has a divergent manner, where both too short-orbital and too long-orbital 
period systems never arrive to Sco X-1 position.
Indeed, none of the above combinations required us to set the initial orbital period of the binary to less than 1.15~days.
In more details, we find two-mode behavior that depends on  whether the initial MT is conservative MT 
(with donors less massive than  $\sim1.5~\msun$) 
or non-conservative MT (donor more massive than $\sim1.5~\msun$). 
If the donor is less massive than about 1.5~$\msun$, the more massive a donor is, 
the weaker the dependence of period at $q=0.3$ is on the initial period.
For donors more massive than 1.5$\msun$ the tendency reverses:  
the higher is the donor mass, the stronger is the dependence of final period on the initial period.
We therefore conclude that we have analyzed the entire initial orbital 
period range that can produce binary systems at the orbital period as in Sco X-1.

\section{Conclusion}
We have conducted detailed simulations of the binary evolution
of Sco X-1. Our simulations show that the commonly
used prescription for magnetic braking, which is based on the observations
of MS stars, is insufficient to  to explain the case of Sco X-1, where the donor is an evolved star (subgiant).
Namely it provides a substantially  lower mass accretion rate than observed, by at least an order of magnitude.

We suggest a different model of magnetic braking, which is
suitable for stars with strong winds,
such as the subgiant donor in Sco X-1. It turns out that this new
model makes possible the formation of a binary system
with the parameters that closely match Sco X-1 from a wide range of systems.
In select models (see Fig.~\ref{fig:tracks_massive_boost_1.42}) with the observed 
period and mass ratio of Sco X-1 we obtain as little as $\approx10\%$ discrepancy between the simulated and observed accretion rate and in all
cases the accretion rate is comparable to the observations. The commonly used prescription gives approximately an order of magnitude lower accretion rate than observed.

Based solely on the known period, mass ratio and accretion rate, we couldn't constrain the parameters of the possible progenitor system.
By varying the initial period we were able to obtain a binary with the observed
period, mass ratio and comparable mass accretion rate from initial NS masses in the range of 1.24 to 1.6~$\msun$ and donor ZAMS masses from 1.0 to 1.6~$\msun$. 
However, based on the maximum effective temperature of the donor, which is 4800~K, the systems with donor ZAMS mass $\ga1.6$~$\msun$ are unlikely to be the progenitors
because in this case even for a 1.3$\msun$ neutron star the resulting effective temperature of the donor exceeds the 4800~K limit.
We find that the effective temperature of the Sco X-1 donor rises with both the donor ZAMS mass and the initial NS mass.
Under the standard assumption that the NS mass is 1.42~$\msun$, the most likely ZAMS mass of the donor is from 1.4 to 1.5~$\msun$, based on the observed accretion rate.

To make sure that the origin of a neutron star (e.g., whether it was formed via electron capture instead of core collapse)
does not affect the theoretically anticipated mass accretion rate, we also tested initially less massive neutron stars, 1.24~$\msun$.
As can be expected from the estimates provided in Section~\ref{sec:mb}, in case of the standard magnetic braking,
the accretion rates at which the binaries with low-mass neutron stars reach the Sco X-1 point (in terms of mass ratio and period)
are much lower than observed in Sco X-1. At the same time, both initially less massive (1.24~$\msun$) and more massive (1.42~$\msun$)
neutron stars have accretion rates comparable to the observations at the Sco X-1 point with wind-boosted magnetic braking prescription.
Thus, although it is not possible to infer from our simulations whether or not Sco X-1 underwent accretion-induced collapse in the past,
if it did, the regular magnetic braking scheme is still not sufficient to explain the observed accretion rate.

Without additional detailed simulations we can't rule out the possibility that a different mass ratio of the Sco X-1
could change the accretion rate anticipated from the standard magnetic braking scheme in a way that the difference between it and the observed value becomes smaller
than an order of magnitude. However, based on the estimates conducted in Section~\ref{sec:mb} we consider that this is unlikely.

Keeping in mind the existing problems with binary simulations of the 
known LMXBs, we anticipate that this new magnetic braking model can (and should)
be used in the simulations of other LMXBs with giant or subgiant donor and
a compact accretor. 
For example, for the majority of LMXBs mentioned in \cite{Podsiadlowski02}, the observed mass accretion
rate is approximately an order of magnitude higher than expected. 
There is also a major mismatch between the observed period decay in A0620-00 and the period
decay expected for this system from the standard magnetic braking prescription. 
The observed decay is $\sim 0.6~\rm{ms\ yr^{-1}}$ \citep{GonzalezHernandez14}. An  estimate
that assumes conservative MT and the standard magnetic braking law with $\gamma = 3$ 
(Equation~\ref{eq:skum}) provides the period decay to be $\sim 0.05~\rm{ms\ yr^{-1}}$ \citep[see also for discussion][]{GonzalezHernandez14}.
Applying the wind-boosted modified magnetic braking 
law we obtain the decay of $0.35~\rm{ms\ yr^{-1}}$, which is closer to the observed value;
$\tau$-boost that increases the surface magnetic field can explain remaining discrepancy.
We note that this system is a candidate circumbinary disk systems 
\citep{2014ApJ...788..184W}, and hence may have an additional mode for the angular momentum loss.

The LMXBs in elliptical galaxies, 
where they are thought to be the main source of X-ray radiation, 
can be observed with quite low detection limits of $\sim10^{36} \rm{erg~s^{-1}}$.
These data can be used in conjunction with population synthesis models 
to examine the formation scenarios of LMXBs.
When an X-ray luminosity function (XLF) of an elliptical galaxy is simulated with population synthesis models,
it turns out to be very sensitive to the magnetic braking prescription used \citep{Fragos08}. 
Depending on the other free parameters of the population synthesis model, 
the discrepancy between
XLFs obtained from different magnetic braking prescriptions reaches an order of magnitude.  
The probability that a simulated XLF is consistent with the observed one also differs drastically for
different magnetic braking prescriptions, from practically zero to comparable to unity, and  
the majority of population synthesis models still gives results that are very unlikely to
be consistent with the observations \citep{Fragos08}. 
With the boosted model for magnetic braking, we also can foresee that not only mass 
accretion rates can be different, but also a class of LMXB systems, 
deemed to be transient with the standard magnetic braking, 
will become persistent.
We anticipate that the new generation of population synthesis models made to 
simulate the XLF functions of elliptical galaxies,
if switched to our modified model of magnetic braking, might give substantially different results.

\section*{Acknowledgments}

KP was supported by Golden Bell Jar Scholarship.
NI thanks NSERC Discovery and Canada Research Chairs Program.
This research has been enabled by the use of computing resources provided
by WestGrid and Compute/Calcul Canada.

\bibliographystyle{mn2e}
\bibliography{mb_bibl}

\end{document}